# Cooling dynamics and thermal interface resistance of glass-embedded metal nanoparticles

Vincent Juvé[1], Mattia Scardamaglia[1], Paolo Maioli[1], Aurélien Crut[1]*, Samy Merabia[2], Laurent Joly[2], Natalia Del Fatti[1] and Fabrice Vallée[1]

[1] *FemtoNanoOptics group, LASIM*
[2] *Theory and Modeling group, LPMCN*
*Université Lyon 1-CNRS,*
*43 Bd du 11 Novembre 1918, 69622 Villeurbanne, France*
*Corresponding author: acrut@lasim.univ-lyon1.fr

**ABSTRACT**

The cooling dynamics of glass-embedded noble metal nanoparticles with diameters ranging from 4 to 26 nm were studied using ultrafast pump-probe spectroscopy. Measurements were performed probing away from the surface plasmon resonance of the nanoparticles to avoid spurious effects due to glass heating around the particle. In these conditions, the time-domain data reflect the cooling kinetics of the nanoparticle. Cooling dynamics are shown to be controlled by both thermal resistance at the nanoparticule-glass interface, and heat diffusion in the glass matrix. Moreover, the interface conductances are deduced from the experiments and found to be correlated to the acoustic impedance mismatch at the metal/glass interface.

I. INTRODUCTION

With the development of nanometric size devices, fundamental understanding and modeling of heat transfer and thermal transport at the nanoscale are now becoming key technological issues. For instance, these processes may constitute important limits in the functioning of nanoelectronic devices, the resistance of nanomaterials under strong excitation, or lead to strong thermal coupling of nanoobjects. Actually, major fundamental differences between thermal transport at macroscopic and nanometric scales are the breaking of the diffusive model and the increased role of interface-mediated effects at the nanoscale. The latter are particularly important in the context of heat dissipation from a nanometer-sized object to its environment, and result in an increasing role of interface thermal resistance ("Kapitza resistance" [1-3]) with size reduction.[4]

Motivated by these fundamental and technological interests, experimental methodologies to study the thermal properties of nanoobjects and nanomaterials are thus rapidly developing. Current approaches include scanning thermal microscopies and time-resolved pump-probe spectroscopy.[4] The former methods use temperature-sensing tips to probe the spatial distribution of temperature,[5-7] while the latter approach is based on following the heat transfer kinetics after excitation of a material formed by a large ensemble of nanoobjects in a solid or liquid matrix. Its principle consists in selectively heating the nanoobjects by a "pump" pulse, and following the dynamics of their subsequent cooling by energy transfer to their environment (Fig. 1). This is done *via* a time-delayed "probe" pulse monitoring an optical property that depends on the temperature of the nanoobjects. Provided that the connection between the measured signal and nanoparticle temperature is known, the kinetics of the temperature decay can thus be determined. As it is ruled by both heat transfer at the particle-matrix interface and heat diffusion in the matrix (Fig. 1), it thus contains information on the former process, i.e., on the Kapitza thermal resistance at the particle-matrix interface.[8-13]

Most previous experiments were carried out in colloidal solutions of metal nanoparticles, and have addressed the impact of nanoparticle size,[8] solvent composition,[9] and interface layer (e.g., using nanoparticles encapsulated in a silica or polymeric shell[11, 13]). In spite of their technological interest, only few experiments were reported on nanoparticles embedded in a solid matrix.[10] However, the limited precision of the data obtained using time-resolved X-ray diffraction[10] precluded a clear separation of the interface and heat diffusion effects in this situation. Actually, only the former effect was considered in the fitting procedure (assuming an exponential decay of the measured signal), which was *a posteriori* justified by the size dependence of the measured cooling times (expected to scale with nanoparticle radius, $R$, or its square, $R^2$, for interface- or diffusion-limited cooling processes, respectively). Such approach can lead to an underestimation of interface conductance, of an amount that depends on nanoparticle size (the interface and diffusion processes being expected to dominate the cooling kinetics in the small and large particle range, respectively).

Here, we report on investigation of the cooling kinetics of noble metal nanospheres embedded in glass using high-sensitivity time-resolved pump-probe experiments. Measurements were performed for a large range of nanoparticle sizes (diameter from 4 to 26 nm) and for different nanoparticle/glass compositions. Special care was devoted to the probing process of nanoparticle cooling, by analyzing the dependence of the measured thermal relaxations on probe wavelength. For all measurements, a non-exponential decay of the measured signal has been observed, signature of a contribution from heat diffusion in the matrix. Full modeling of the experimental data permits extraction of the interface conductance and estimation of its dependence on metal and glass composition.

## II. SAMPLES OF METAL NANOPARTICLES-DOPED GLASSES

Experiments were performed with two different sets of samples formed by silver, gold or silver-gold quasi-spherical nanoparticles embedded in different glass matrices. This permits to test the fitting procedure of the time-resolved experimental data, and to estimate the impact of glass and nanoparticle composition on the interface resistance. All samples were synthesized using a fusion and heat treatment technique. The first set of samples is formed by either monometallic (gold) or bimetallic (gold-silver) nanoparticles in a crystal glass ("glass 1": 53% $SiO_2$, 30% $PbO$, 12% $K_2O$, 2% $Na_2O$, 2% $Sb_2O_5$, 1% $As_2O_3$). The second set is formed by silver nanospheres embedded in a 50% $BaO$, 50% $P_2O_5$ glass ("glass 2"). For both types of samples, the average size of the nanoparticles has been either directly determined by transmission electron microscopy,[14] or deduced from the period of their acoustic vibrations measured by Raman and/or time-resolved spectroscopies.[15, 16] The average nanoparticle diameter is 9 nm for the gold particles in glass 1, ranges from 8.9 to 11.2 nm for the bimetallic particle/glass 1 samples and from 4.2 to 26 nm for the silver/glass 2 samples. The optical spectra of all the samples show enhanced absorption due to the localized surface plasmon resonance (SPR) of the metal nanoparticle, around 420 nm (Fig. 2(a), inset) and 530 nm for the silver and gold samples, respectively. A well-defined SPR is also observed for the bimetallic particle samples, suggesting alloying of gold and silver[17] and allowing the deduction of alloy stoichiometry from SPR position.[17, 18]

The thermal properties of glass matrices determine the contribution of heat diffusion to the measured kinetics. For glass 1, the specific heats tabulated in SciGlass database range from 1.60 to 1.81 $J.m^{-3}.K^{-1}$, and thermal conductivities from 0.8 to 0.9 $W.m^{-1}.K^{-1}$. The former are in the 1.91 to 2.16 $J.m^{-3}.K^{-1}$ range for glass 2, while its thermal conductivity is not tabulated. However, it is expected to lie below 0.4 $W.m^{-1}.K^{-1}$, considering the thermal conductivity of the two components of this glass (2.5 $10^{-3}$ $W.m^{-1}.K^{-1}$ for $P_2O_5$, and in the 0.21 - 0.43 $W.m^{-1}.K^{-1}$ range for BaO).

## III. TIME-RESOLVED EXPERIMENTAL SET-UP

Time-resolved experiments were performed using a standard two-color pump-probe setup, based on a home-made Ti:sapphire oscillator delivering pulses of about 20 fs at 890 nm with a repetition rate of 76 MHz. The output pulse train was split into two parts, one being frequency-doubled to 445 nm in a 500 µm thick BBO crystal. The two fundamental and harmonic beams were focused on the samples using two different lenses. One beam is used to selectively heat the metal nanoparticles (pump pulse), while the second one monitors the time-dependent transmission changes $\Delta Tr$ of the sample (probe pulse) induced by the pump beam. Experiments were performed using either the near-infrared (890 nm) or blue (445 nm) beam as the pump beam (the probe beam being then in the blue or near-infrared, respectively). Under our experimental conditions, the maximum increase of the lattice temperature $\Delta T_0$ of the nanoparticle is about 30 K (1 K) when pumping at 445 nm (890 nm). For both probe wavelengths $\lambda_{pr}$, change of the sample reflectivity can be neglected,[19] so that the measured normalized transmission change ΔTr/Tr can be identified with the change of sample absorption:

$$\frac{\Delta Tr}{Tr}(\lambda_{pr}) = -\Delta\alpha(\lambda_{pr})\,L \qquad (1)$$

where L is the sample thickness, and α its absorption coefficient.

The time delay between pump and probe pulses was varied using a mechanical delay stage. High sensitivity detection of the pump-induced changes of the probe pulse transmission was achieved by mechanical chopping of the pump beam at 100 kHz, combined with a synchronous and differential detection.

## IV. EXPERIMENTAL RESULTS

The transient transmission change ΔTr/Tr measured for 26 nm diameter Ag particles embedded in glass 2 is illustrated in Fig. 2, for probe wavelengths of 445 nm and 890 nm, i.e., when probing is performed close (Fig. 2(a)) or away (Fig. 2(b)) from the SPR. As expected, in spite of a smaller heating of the particles by the pump pulse, the ΔTr/Tr amplitude is much larger in the former probing condition (Fig. 2(a)), due to enhancement of the optical response in vicinity of the SPR.[20, 21] In both cases, short delay signals (t ≤ 5 ps) reflect relaxation of the photoexcited electrons and thermalization of the electrons and lattice in each particle at temperature $T_p$ (i.e., internal thermalization). They are followed by weak oscillations due to the coherent acoustic vibrations of the nanoparticles, over about 30 ps. These two processes have been extensively studied and modelled,[15, 16, 19, 22-24] and will not be further discussed here, where we will focus on the long-delay ΔTr/Tr decay over a few hundred picoseconds time scale, which contains information on the cooling of the nanoparticles induced by energy exchanges with their environment.

Extraction of this information requires connecting the measured transient change of the sample optical properties to nanoparticle temperature $T_p$. In most previous optical pump-probe experiments, a direct proportionality between changes in transmittance and temperature has been implicitly assumed. Such simple assumption cannot be performed when probing close to a relatively narrow SPR, as in the case of silver nanoparticles. This is illustrated by the probe wavelength dependence of the long-delay ΔTr/Tr signals (Fig. 2(c)), isolated from bare signals (Fig. 2(a) and 2(b)) by substraction of the short-time contributions due to internal thermalization and acoustic vibrations of the particles. The origin of this ΔTr/Tr (i.e., Δα, see Eq. (1)) dependence can be identified by analyzing the different contributions to the absorption of nanoparticle samples. For the relatively small particles investigated here, absorption can be written using the Mie theory in the dipolar approximation:[25]

$$\alpha(\lambda) = \frac{18\pi p}{\lambda} \frac{\varepsilon_m(\lambda)^{3/2} \varepsilon_2(\lambda)}{[\varepsilon_1(\lambda) + 2\varepsilon_m(\lambda)]^2 + \varepsilon_2(\lambda)^2} \quad (2)$$

where p is the particle volume fraction (typically $10^{-4}$) and λ the wavelength. In this expression, both the complex dielectric constant of the metal composing the nanoparticles, $\varepsilon=\varepsilon_1+i\varepsilon_2$, and the real one of the surrounding glass matrix $\varepsilon_m$ are modified in the long time-scale measurements considered here. For probe delays longer than about 5 ps, the change of ε is associated to the rise of the temperature $T_p$ of the internally thermalized nanoparticle, and is proportional to it when probing away from the interband transitions of the metal.[21, 22] Additionally, cooling of a nanoparticle by energy transfer to a surrounding matrix of finite thermal conductivity leads to a local rise of matrix temperature, and thus to a modification of $\varepsilon_m$. This essentially translates into a shift of the SPR wavelength (Eq. (2)), and thus into a modification of the sample absorption $\alpha$. The signal measured in time-resolved experiments thus *a priori* contains contributions from both nanoparticle cooling and glass heating kinetics.

As the SPR wavelength of a nanoparticle is only sensitive to the local dielectric constant of its environment over a distance of the order of its radius,[26-28] the amplitude of this effect can be estimated by assuming a local mean temperature $T_m$ of the glass surrounding a particle. Assuming a weak modification of the system properties, the change of the sample absorption $\Delta\alpha$ can thus be connected to the rises of the temperature of the nanoparticle and glass:

$$\Delta\alpha(\lambda) = \left( \frac{\partial\alpha}{\partial\varepsilon_1}\frac{d\varepsilon_1}{dT_p} + \frac{\partial\alpha}{\partial\varepsilon_2}\frac{d\varepsilon_2}{dT_p} \right)_\lambda \Delta T_p + \left( \frac{\partial\alpha}{\partial\varepsilon_m}\frac{d\varepsilon_m}{dT_m} \right)_\lambda \Delta T_m . \qquad (3)$$

The first term dominates the short time delay response and has been extensively discussed in the context of the investigation of the ultrafast response and electron cooling kinetics of metal nanoparticle.[19] The probe wavelength dependence of the second one is estimated using Eq. (2) and the tabulated dielectric constants of silver and gold,[29] taking into account that the temperature dependence of the dielectric constant of glasses is almost wavelength independent (with typically $d\varepsilon_m/dT_m \approx 10^{-5} K^{-1}$). To compute the maximal possible contribution of this effect, as a crude approximation the local rise of the glass temperature $\Delta T_m$ was identified with the maximum induced temperature rise of the nanoparticle lattice, i.e., about 30 K and 1 K for a pump wavelength of 445 nm and 890 nm, respectively. The estimated glass contribution to the sample transmission change $\Delta Tr/Tr = -\Delta\alpha L$ (second term in Eq. (3)) is about $3\times10^{-5}$ when probing close to the SPR, at $\lambda_{pr} = 445$ nm, a value comparable to the experimentally observed $\Delta Tr/Tr$ (the thermal signal extracted from Fig. 2(a) presents a maximum of about $7\times10^{-5}$). This suggests that local glass heating significantly contributes to the measured transient transmission change of the sample when probing in the vicinity of SPR. Conversely, the term related to glass heating is expected to decrease by about two orders of magnitude when shifting the probe wavelength from the blue (445 nm) to near-infrared (890 nm) part of the spectrum, this result being a consequence of reduction of the $\Delta\alpha$ sensitivity on the glass dielectric constant away from SPR. In this case, its contribution to $\Delta Tr/Tr$ (of the order of a few $10^{-7}$) is negligible compared to the experimentally measured $\Delta Tr/Tr$ (the thermal signal extracted from Fig. 2(a) presents a maximum close to $10^{-5}$). Therefore, the contribution of glass heating to the time-resolved signals can be neglected away from SPR, i.e., in the near-infrared, the experimental signals being then proportional to nanoparticle temperature rise.

## V. DISCUSSION

The rate at which heat dissipates from a nanoparticle depends both on the thermal interface resistance which governs energy transfer at the interface between the nanoparticle and its surrounding, and on heat diffusion in the surrounding medium (Fig. 1). As in some previous works involving colloidal solutions,[9, 12] modeling of the cooling kinetics of our glass-embedded metal nanoparticles was performed taking into account both effects. The electron and lattice temperatures in a nanoparticle have been assumed to be well-defined and identical (to $T_p$). As absorption of the pump pulse initially results in the creation of an athermal electron distribution out of equilibrium with the lattice, this assumption in only valid after internal energy redistribution in a particle, i.e., after a few picoseconds.[19] The nonequilibrium pump-probe approach also raises the question of energy redistribution among the lattice modes, i.e., proper definition of the lattice temperature, or equivalently, the possibility of hot-phonon effects on the studied time scale. This effect can not only influence the electron-lattice thermalization kinetics inside a nanoparticle, but also energy transfer to its surrounding via coupling of the nanoparticle and matrix vibrational modes. In particular, it is interesting to point out that the damping time of the fundamental radial mode of Ag particles in glass 2, determined in previous experiments,[15, 30] is much faster than the cooling time of the nanoparticles measured here. This suggests different energy transfer rates of the vibrational modes of the particles to the matrix, and a possible impact of the energy redistribution among the nanoparticle vibrational modes on the observed global energy losses. However, these processes are relevant when quantitatively comparing the computed and experimental interface resistance, and investigating the elementary mechanisms at its origin,[31, 32] which is out of the scope of this paper. Here, we will assume that the nanoparticle temperature can be defined throughout the cooling process (i.e., that all internal thermalization processes are fast on the timescale of the nanoparticle cooling). Temperature will also be assumed to be uniform over the nanoparticle, which is justified by the high thermal conductivity of metals. The temperature $T_m$ of the glass matrix around a particle is assumed to depend only on the distance from the particle centre (since the volumic fraction of nanoparticles is of the order of 10$^{-4}$, the samples are sufficiently dilute to assume that the particles are independent, i.e., matrix heating by other particles is neglected).

Heat dissipation from a spherical nanoparticle of radius $R$ is then governed by a set of two equations describing heat flux at the particle-matrix interface (Eq. (4)) and heat diffusion within the glass matrix (Eq. (5)):

$$\frac{\partial T_p(t)}{\partial t} = -\frac{3G}{R c_p}\left(T_p(t) - T_m(R,t)\right), \tag{4}$$

$$c_m \frac{\partial T_m(r,t)}{\partial t} = \Lambda_m \frac{1}{r}\frac{\partial^2}{\partial r^2}\left(r T_m(r,t)\right). \tag{5}$$

where $c_{p(m)}$ is the particle (matrix) specific heat per unit volume, $\Lambda_m$ the thermal conductivity of the matrix, $G$ the interface thermal conductance, and $r$ the distance to the particle centre. Operating in the Laplace domain, one obtains the following expression for the time-dependence of the particle temperature:[33, 34]

$$\Delta T_p(t) = \frac{2kR^2 g^2 \Delta T_0}{\pi} \int_0^{+\infty} \frac{du\, u^2 \exp(-\kappa u^2 t / R^2)}{\left[u^2(1+Rg) - kRg\right]^2 + (u^3 - kRgu)^2} \tag{6}$$

where $\Delta T_0$ is the initial temperature increase of the particle, $\kappa=\Lambda_m/c_m$, $k=3c_m/c_p$ and $g=G/\Lambda_m$.

If one of the involved mechanisms, i.e., interface-resistance or heat-diffusion, limits the nanoparticle cooling kinetics, a much simpler expression is obtained. A mono-exponential (Eq. (4)) or non-exponential (Eq. (5)) decay of the nanoparticle excess temperature $\Delta T_p$ is then expected, respectively. Such approximation has been frequently performed in time-resolved studies of nanoparticle cooling, considering either only interface[10] or diffusion effects.[8, 11, 13]

The validity of this assumption was *a posteriori* discussed based on the difference dependence of the cooling times on nanoparticle size, which are proportional to R and $R^2$ for interface- and diffusion-limited processes, respectively. However, this approach is limited to particle size and environment conditions, for which one of the process limits the cooling kinetics, i.e., is much slower than the other. This is not the case in our experimental conditions, where both mechanisms have similar timescales. Their relative amplitudes reflect in the concavity of the experimental relaxation traces when shown on a semi-log plot (Fig. 2(c) and Fig. 3). Indeed, a larger concavity reflects a larger heat diffusion contribution, permitting to quantify the interface and heat diffusion contribution, provided experimental signals display a high enough signal-to-noise ratio.

Experimental data were reproduced assuming that the measured transient transmission change $\Delta Tr/Tr$ is proportional to the nanoparticle temperature decay $\Delta T_p$ computed using the full thermal model.[9, 12] In this comparison, $\Delta T_p$ is numerically calculated using Eq. (6) and the thermal constants tabulated for noble metals (specific heats $2.5\ 10^6$ and $2.4\ 10^6$ $J.m^{-3}.K^{-1}$ for gold and silver, respectively) and for the glass matrix (section II). The interface conductance G is used as a parameter, together with $\Lambda_m$ when the latter is not precisely known (section II), a $Chi^2$ minimization procedure comparing the theoretical and experimental data being then used to extract G and $\Lambda_m$. This is illustrated in Fig. 2(c), showing an excellent reproduction of the long time-delay data measured in 26 nm silver nanoparticles in glass 2 using an interface conductance G = 315 $MW.m^{-2}.K^{-1}$ and a glass heat conductivity $\Lambda_m$ = 0.21 $W.m^{-1}.K^{-1}$, in the range expected for a $BaO-P_2O_5$ glass ( < 0.4 $W.m^{-1}.K^{-1}$). Note that in contrast, fitting the experiments carried out probing close to the SPR, i.e., at $\lambda_{pr}$ = 445 nm, using the same approach leads to an unrealistically large value of the glass heat conductivity (0.97 W.m$^{-1}$.K$^{-1}$), confirming that the $\Delta Tr/Tr$ time-dependence does not directly reflect the decay of nanoparticle temperature in this case. In the following, we will thus focus on the data obtained with near-infrared probing.

For all the investigated samples, with nanoparticle mean diameter ranging from 4 to 26 nm, both interface effects and glass heat diffusion have been found to significantly contribute to the nanoparticle cooling kinetics. This is illustrated in Fig. 3 showing the decay of the long-delay $\Delta Tr/Tr$ signal measured for two samples with different nanoparticle size and composition and for two different glasses (9 nm gold or 26 nm silver nanoparticles embedded in glass 1 or 2, respectively). The experimental decays cannot be reproduced assuming one limiting mechanism (a fitting attempt only taking into account interface effect is presented in Fig. 3). Moreover, the thermal conductivity $\Lambda_m$ of glass 1 derived from the full fitting procedure (0.9 W.m$^{-1}$.K$^{-1}$) is close to the tabulated values (0.8-0.9 W.m$^{-1}$.K$^{-1}$), confirming the validity of our approach. In the following, experiments performed on nanoparticles embedded in glass 1 were reproduced using a fixed value of $\Lambda_m$ = 0.9 W.m$^{-1}$.K$^{-1}$. In the case of BaO-P$_2$O$_5$ (glass 2), $\Lambda_m$ was left as a free parameter. Its value, deduced from the fitting procedure, was seen to always fall below 0.4 W.m$^{-1}$.K$^{-1}$, as expected (see section II).

The interface resistances 1/G estimated for the different samples are displayed in Fig. 4. In contrast to glass 1 samples, Ag/glass 2 samples show significant variations of the value deduced for the interface resistance, in the 2.5 to 5 GW$^{-1}$.m$^2$.K range (Fig. 4), probably due to slightly different synthesis conditions between samples.

Nevertheless, systematic variations were observed as a function of the materials involved in the interfaces, i.e., gold/glass 1, gold-silver alloys/glass 1 and silver/glass 2, clearly showing a dependence of interface resistance on the composition of the nanoparticle and its environment (Fig. 4). To quantify this variation, we have characterized the interfaces by the acoustic impedance mismatch of the particle and matrix materials, $Z_p/Z_m$, which controls the interface resistance in the "acoustic mismatch" model.[4] A good correlation is obtained between the estimated interface resistances and the $Z_p/Z_m$ values computed using the tabulated acoustic impedances of gold ($Z_p$ = 63 10$^6$ kg.m$^{-2}$.s$^{-1}$), silver ($Z_p$ = 38 10$^6$ kg.m$^{-2}$.s$^{-1}$), and the ones measured for glass 1 ($Z_m$ = 14.9 10$^6$ kg.m$^{-2}$.s$^{-1}$) and glass 2 ($Z_m$ = 17.6 10$^6$ kg.m$^{-2}$.s$^{-1}$).[30] This correlation is consistent with the measured dependence of the damping of the fundamental acoustic mode of metal nanoparticles, due to transfer of their energy to the matrix, i.e., corresponding to a specific particle-matrix energy transfer channel.[15, 30]

Though this correlation is fully consistent with the "acoustic mismatch" model, we emphasize here that the measured interface resistances not only depend on the nature of the materials, but also on the quality of their interface. Time-resolved investigation of the breathing modes of silver nanoparticles in glass showed measured damping times slightly larger than computed ones, which was attributed to a non-perfect nanoparticle-glass contact.[30] Such variations of the nanoparticle/matrix contact condition may be responsible for the dispersion of thermal conductances measured for silver/glass 2 samples. Additional more systematic studies with, in particular, a better control of the nanoparticle-matrix contact, are necessary to confirm the correlation between the interface conductance and acoustic mismatch suggested by our results.

## VI. CONCLUSIONS

Using time-resolved two-color pump-probe spectroscopy, we have analyzed the relaxation kinetics of noble-metal nanoparticles of sizes ranging from 4 to 26 nm embedded in two different glasses. The measured time-resolved signals have been shown to reflect changes in nanoparticle temperature only when probing away from an optical resonance of the material, i.e., SPR. This conclusion is supported by the fact that fitting the experimental data obtained in the non-resonant condition yields glass thermal conductivity in excellent agreement with the tabulated ones, in contrast to the data obtained for resonant probing.

The experimental data were reproduced including both interface resistance effects and heat diffusion in glass. This approach permits extraction of the thermal interface resistance and its investigation as a function of the nature of the materials forming the interface. Our results suggest correlation between the interface resistance and the nanoparticle-glass acoustic mismatch. Further investigation in samples with better controlled nanoparticle/glass contacts are required to confirm these results. This work also raised the question of the elementary mechanisms involved in the measured cooling kinetics. In particular, elucidation of the interplay between energy transfer between the individual vibrational modes of a nanoparticle and its surrounding matrix and the energy redistribution mechanism between the different modes of one material would be particularly interesting. Systematic studies of the parameters influencing solid-solid interface resistances at a nanometric scale would also be of large fundamental and technological interest.

ACKNOWLEDGEMENTS

The authors thank A. Mermet and E. Duval for providing the Au and AuAg/glass 1 samples, and S. Omi for providing silver/glass 2 samples, and J.-L. Barrat for useful discussions. This work was funded by the "Opthermal" grant of the Agence Nationale de la Recherche. NDF acknowledges support from the Institut Universitaire de France (IUF).

Figures

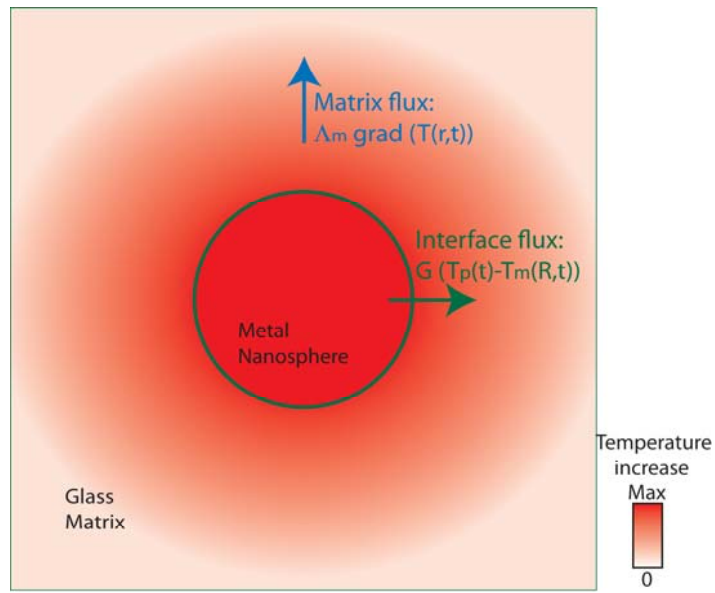

**FIG. 1.** Schematics of the cooling dynamics of glass-embedded spherical nanospheres after ultrafast heating by a femtosecond pump pulse. Cooling requires heat transfer at the metal/glass interface (controlled by the interface conductance G), and heat diffusion in the matrix (governed by the glass matrix thermal conductivity $\Lambda_m$).

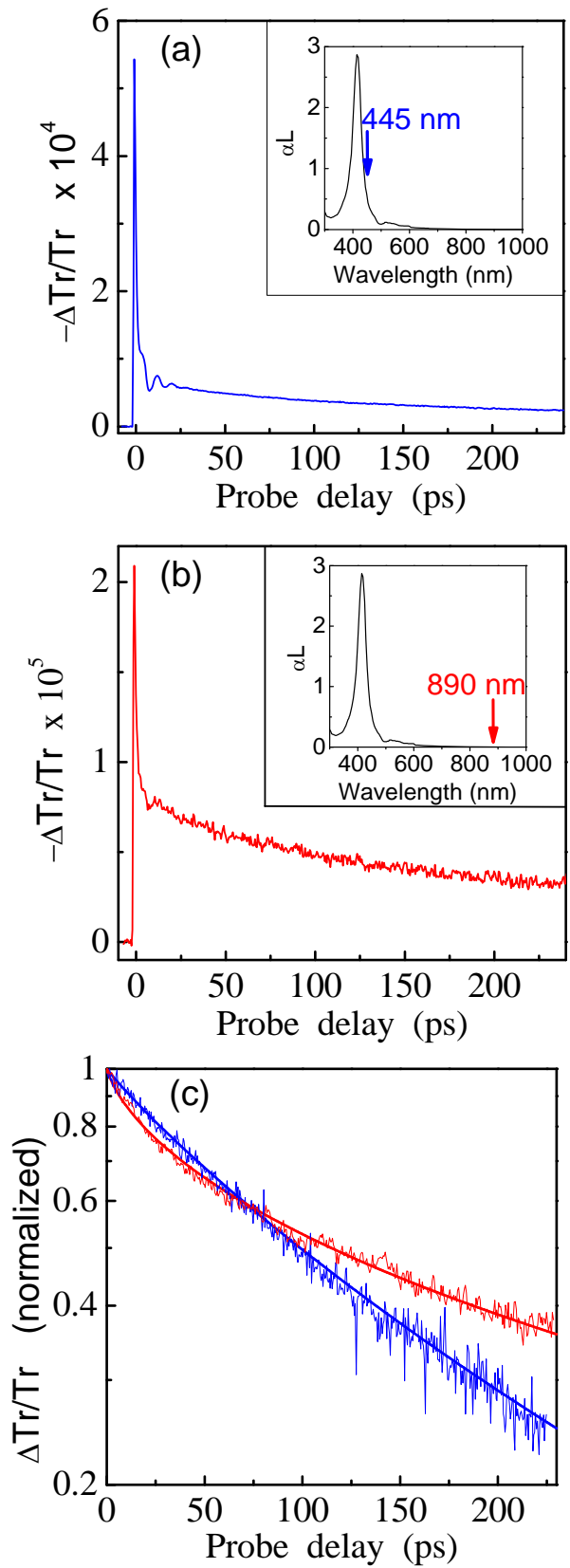

FIG. 2. (a) and (b) Time-dependent transmission change ΔTr/Tr measured in ultrafast pump-probe experiments performed with 26 nm diameter silver nanoparticles embedded in glass 2. The pump and probe wavelengths are 890 nm and 445 nm (a), or 445 nm and 890 nm (b), respectively. Insets present the position of the probing wavelength on the absorption spectrum (red and blue arrows, respectively). (c) Normalized long-delay signals corresponding to (a) and (b) blue and red plots, respectively. Full lines are fits using Eq. (6) with G=100 MW.m$^{-2}$.K$^{-1}$ and $\Lambda_m$=0.97 W.m$^{-1}$.K$^{-1}$ (blue line) and G=315 MW.m$^{-2}$.K$^{-1}$ and $\Lambda_m$=0.21 W.m$^{-1}$.K$^{-1}$ (red line).

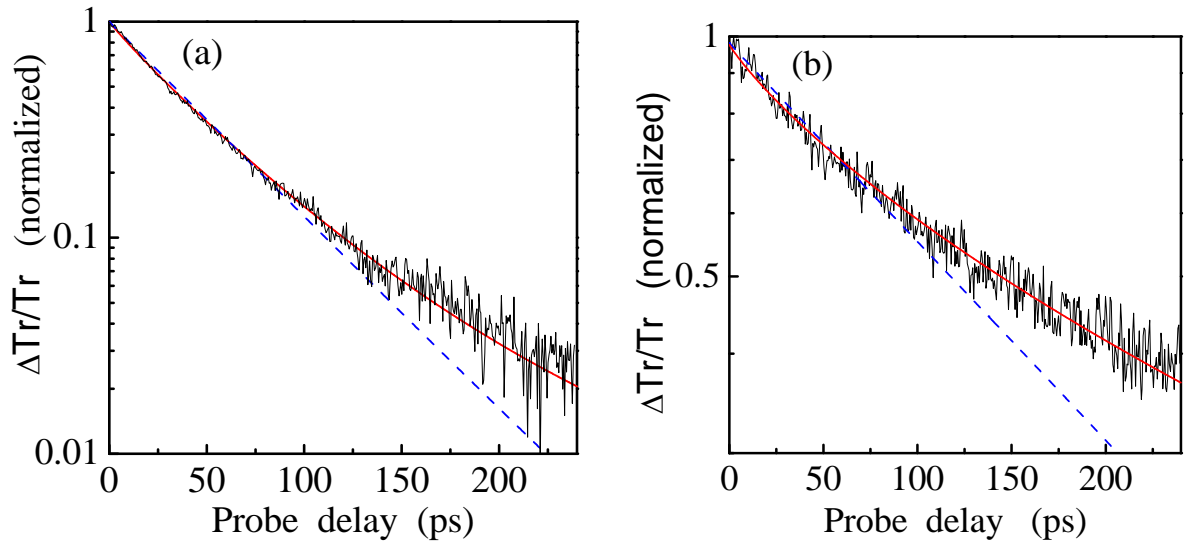

**FIG. 3.** Time dependence of the normalized transmission change ΔTr/Tr measured in 9 nm gold nanoparticles embedded in glass 1 (a) and 26 nm silver nanoparticles embedded in glass 2 (b). The probe wavelength is 890 nm. The full red lines correspond to fits including both interface thermal resistance and heat diffusion effects (Eq. (6)), and the dashed blue lines to fits including only interface thermal resistance (Eq. (4)).

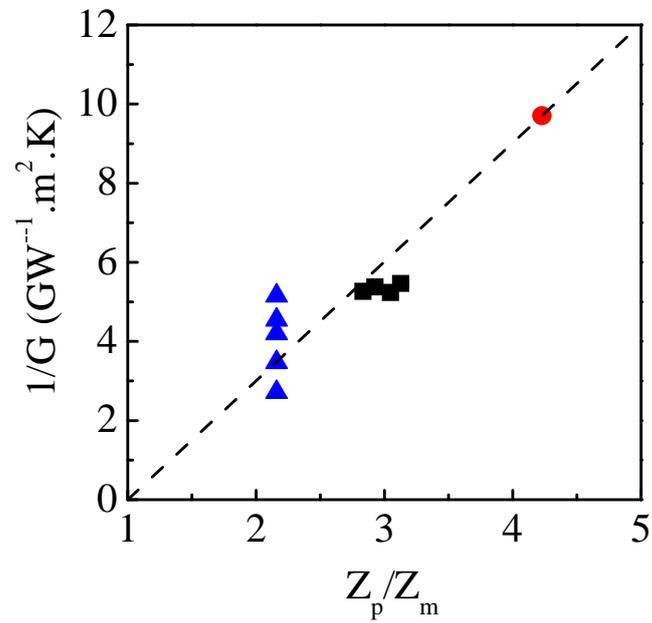

**FIG. 4**. Interface thermal resistance 1/G measured in the different nanoparticle samples as a function of the acoustic impedance mismatch of the nanoparticle and matrix materials $Z_p/Z_m$. Red circle: Au/glass 1 sample (nanoparticule diameter: 9 nm); black squares: AuAg/glass 1 samples (nanoparticule diameter/gold fraction, from left to right: 16.1 nm/17%, 19 nm/23%, 22 nm/30% and 20.5 nm/35%); blue triangles: Ag/glass 2 samples (nanoparticule diameter, from bottom to top: 9, 24, 9.8, 4.2 and 26 nm). The dashed line is a guide for the eye.